\documentstyle[psfig,sprocl]{article}

\def\be{\begin{equation}}
\def\ee{\end{equation}}
\def\bea{\begin{eqnarray}}
\def\eea{\end{eqnarray}}

\def \susy {supersymmetry }

\def\L {\Lambda }
\def\l {\lambda }

\def \g {\gamma }

\def \s {\sigma }
\def \e {\epsilon }
\def \ud {{1 \over 2} }

\def \bea {\begin{equation} }
\def \eea {\end{equation} }

\def \Eslash {E \kern-.5em\slash }
\def \pslash {p \kern-.5em\slash }
\def \kslash {k \kern-.5em\slash }

\begin{document}

\title{R-PARITY VIOLATING CONTRIBUTIONS TO FLAVOR CHANGING AND CP
VIOLATION EFFECTS IN FERMION AND SFERMION PAIR PRODUCTION}

\author{ M. CHEMTOB and G. MOREAU }

\address{Laboratoire de la Direction des Sciences de la Mati\`ere \\
du Commissariat \`a l'Energie Atomique \\ Service de Physique
Th\'eorique \\ CE-Saclay F-91191 Gif-sur-Yvette, Cedex FRANCE}

\maketitle\abstracts{ We examine the contributions from the R-parity
odd interactions, $ \l_{ijk} L_iL_jE_k^c $ and $ \l ' _{ijk}  L_i Q_j
D^c_k$, to the total rates and the CP asymmetry rates in the
production of fermion-antifermion (lepton, down and up quark) pairs
and slepton-antislepton pairs of different flavors at leptonic
colliders.  For the top-charm associated production case, we evaluate
dynamical distributions for the semileptonic top decay events and
estimate the sensitivity reach on the relevant parameters.}

\section{Introduction}

The flavor non-diagonal and/or CP violating effects induced through
the R parity violating (RPV) interactions may become observably
accessible at the high energy leptonic colliders. For the two-body
reactions of fermion-antifermion pair production, $l^-+l^+\to f_J
+\bar f_{J'} ,$ where the produced fermions are charged leptons,
down-quarks or up-quarks of different flavors, $J\ne J'$, the RPV
interactions contribute at the tree level to the rates, $ \s
_{JJ'}$. Furthermore, the differences of rates for the pairs of
CP-conjugate processes, $ M_{JJ'} = l^-+l^+\to f_J +\bar f_{J'},$ and
$ \bar M_{JJ'} = l^-+l^+\to \bar f_J +f_{J'},$ as described by the
asymmetry, ${\cal A}_{JJ'}= (\vert M^{JJ'} \vert ^2- \vert \bar
M^{JJ'} \vert ^2 ) /( \vert M^{JJ'} \vert ^2+ \vert \bar M^{JJ'} \vert
^2) ,$ provide simple CP-odd observables involving interference terms
of tree and one-loop level amplitudes.  The particular mechanism we
study uses the possibility where a complex CP-odd phase is embodied in
the RPV coupling constants, as is allowed thanks to the non trivial structure of the RPV
coupling constants in the quarks and leptons generation spaces. 
One may, of course, think of other RPV induced mechanisms 
in which  the CP-odd complex phase  arises  from the regular interactions in
the  minimal supersymmetric standard model (MSSM).
 It is worth recalling here that the standard model contributions to
the flavor changing rates and/or CP asymmetries for fermion pair
production arise first at the one-loop order and are found to be
exceedingly small.  By contrast, the prospects are on the optimistic
side in most physics extensions beyond the standard model, including the 
MSSM, where the contributions may be bounded by
postulating either a degeneracy of the soft \susy breaking scalars
masses or an alignement of the fermion and scalar superpartners mass
matrices.

  If the initial energy at linear colliders is
sufficient for the production of a pair of sfermions, the analogous
two-body reactions involving the production of slepton pairs,
$l^-+l^+\to \tilde e_{HJ} +\tilde e^\star _{H'J'}, \ [J\ne J',\
H,H'=(L,R) ]$, could also be observed. 
Flavor non-diagonal  sfermion pair production can proceed in the MSSM 
via a  flavor oscillation mechanism.

  In this report, we present a summary of results for the
rates $ \s _{JJ'} $ and the CP rate asymmetries, ${\cal A}_{JJ'} = (
\sigma _{JJ'}- \sigma _{J'J}) /(\sigma _{JJ'}+ \sigma _{J'J} ) $,
obtained by us in studies of the RPV interactions contributions to
fermion pair production, \cite{Chemfer} and to sfermion pair
production.  \cite{Chemsca} We also summarize results from a detailed
examination of the semileptonic top decay signal for single top
production. \cite{topdist}

\section{Fermion pair production}

The flavor non-diagonal fermion pair production process, $l^-+l^+\to
f_J +\bar f_{J'} $, proceeds at the tree level through sfermion
exchange diagrams (t-channel for lepton and down-quark, u-channel for
up-quark and s-channel for lepton and down-quark) controlled by pairs
of distinct RPV interactions coupling constants. Restricted  sets  of  
chirality configurations for the initial and final state fermions are 
generally selected.   The vertex and box
diagrams arising at the one-loop level are controlled by different
pairs of RPV coupling constants. We shall restrict consideration
to the vertex corrections in the $\g -$ and $Z-$ boson exchange. The
quadratic products of the relevant RPV coupling constants, along with
the intermediate fermion and scalar particles, present in the tree and
one-loop amplitudes are displayed in Table \ref{table1}.  This clearly
demonstrates that a large number of the RPV coupling constants can be
probed through the study of these reactions.
\begin{table}
\caption{RPV coupling constants products entering the 
t- and u-channel exchange  amplitudes for fermion-antifermion 
pair production. The column entries refer to the final states with charged
lepton, down-quark and up-quark pairs, respectively, and the two line
entries to the tree and one-loop  level contributions. }
\begin{tabular}{ccc}
\\ 
\hline $ e^-_J e^+_{J'} $ & $ d_J\bar d_{J'} $ & $ u_J\bar u_{J'} $ \\
\hline  \\ 
$\l_{iJ1}\l ^\star _{iJ'1},\ \tilde \nu_{iL}$; $ \l_{i1J}
\l^\star _{i1J'},\ \tilde \nu _{iL}$ & $\l ' _{1jJ}\l ^{'\star }
_{1jJ'}, \ \tilde u_{jL} $ & $\l ^{'\star }_{1Jk}\l ' _{1J'k}, \
\tilde d_{kR} $ \\
\hline \\   $ \l ^{'\star }_{Jjk} \l '_{J'jk}; \
 \l ^{\star }_{iJk} \l _{iJ'k}; \ \l_{ijJ} \l ^{\star }_{ijJ'} $ 
& $\l ^{'\star }_{iJk} \l ' _{iJ'k}$; $\l ^{'\star }_{ijJ'} \l ' _{ijJ} $ &
$\l ^{'\star }_{iJk} \l ' _{iJ'k}$ \\  $ {d_k \choose \tilde u^\star
_{jL} } , \ {u _j^c \choose \tilde d_{kR}}  ;\  \ {e_k \choose \tilde \nu^\star _{iL} };
$ & $ {d_k \choose \tilde \nu^\star _{iL}} , \
{\nu^c_i \choose \tilde d_{kR}} $; & $ {d_k \choose \tilde e^\star _{iL}} ,
\ {e^c_i \choose \tilde d_{kR}} $ \\ 
$  {e_j \choose {\tilde \nu
}_{iL} } , \ {\nu _i \choose \tilde e_{jL}} $   & ${d_j \choose \tilde \nu_{iL}} , \
{u_j \choose \tilde e_{iL}}  , {\nu_i \choose
\tilde d_{jL}} , \ {e_i \choose \tilde u_{jL}}  $  & \\
\hline
\end{tabular}
\label{table1}
\end{table}

Proceeding to the numerical results, we find that the predicted flavor
non-diagonal rates slowly decrease with the center of mass energy and
scale with the the exchanged superpartner mass parameter, $ \tilde m$,
approximately as, \cite{Chemfer} \hfill  $ \s_{JJ'}\approx ({\L \L \over
0.01})^2 ({100 GeV \over \tilde m})^{ p} (1 \ - \ 10) fbarns$, where,
$ \L = \l $ or $\l '$, as the case may be.  The exponent controlling
the superpartner mass dependence lies in the range, $ p \simeq ( 2 \ -
\ 3)$, and decreases with increasing center of mass energy. 
For realistic experimental conditions corresponding to a maximum
number of signal events around, $ N_S =\s {\cal L} =10$, and an integrated
luminosity around, $  {\cal L} = 100 fb^{-1}$, the sensitivity on the 
relevant RPV coupling constants products, as quoted in Table \ref{table1}, 
reads:  $\L \L  < 3 \ 10^{-3} ({\tilde m\over 100 GeV })^{-p/2} 
({ N_S \over 10 })^{1/2 } ({100 fb^{-1} /{\cal L} })^{1/2 }.$ The majority of 
the current indirect bounds on the relevant products, $ \l \l , \  \l ' \l '$,
lie close to the above quoted value, while the  strongest bounds on quadratic products,
such as those arising from $ \mu \to 3l, \ K \to \pi \nu \bar \nu,  \cdots  $
can be evaded  thanks to the existence of several different 
contributions to the fermion pair production amplitudes. An analysis of the
current single coupling constants bounds yields the following list 
of least constrained products of couplings and the corresponding
preferred fermion pair production.   ${\bf  e \bar \tau }: \l_{211} \l_{231,213}
< O(10^{-3} ) , \  \l_{311} \l_{313} < O(10^{-3} )   ; {\bf  \ \mu  \bar \tau }: 
\l_{121} \l_{131} < O(10^{-3} ) , \ \l_{212} \l_{213}< O(10^{-3} ) , 
\l_{312} \l_{313}< O(10^{-3} ) , \  \l_{311} \l_{323}< O(10^{-3} ), 
\  \l_{211} \l_{232}< O(10^{-3} )   ;
  \ {\bf d\bar b}: \l '_{121} \l '_{123} < O(10^{-3} ) ;\ 
 {\bf s\bar b}: \l '_{122} \l '_{123} < O(10^{-2} ) ; 
 {\bf c \bar t}: \l '_{122} \l '_{132} < O(10^{-2} ), \ \l '_{121} \l '_{131} < O(10^{-3} ) .$

We choose to assign the CP-odd phase in such a way that the RPV coupling
constants product in the tree amplitude is real while that of the loop
amplitude has a non-vanishing complex phase, $ e^{i\psi }$. Assuming
that all the relevant RPV coupling constants are set at equal
values,  we find asymmetries of order, $A_{JJ'} \approx
(10^{-2}-10^{-3}) \sin \psi $.  
 The rational dependence of the CP asymmetries on the coupling constants,
$Im((\l \l^*)_{loop} (\l \l^*)_{tree} /(\l ^4)_{tree}$, may lead to
strong enhancement or suppression factors on the CP rate asymmetries
to the extent that the quarks or leptons generation dependence of the
RPV coupling constants presents a hierarchical structure.
Setting tentatively the RPV coupling constants at their current bounds, 
one predicts for $e \bar \tau , \  Im (\l ^{'\star } _{132} \l '_{332} /\l^{\star } _{311} \l_{321}) 
\sim O(10), $ and for, $  t \bar  c :  Im (\l ^{'\star }_{32k} \l '_{33k} /\l ^{'\star } _{12k} \l '_ {13k})   
\sim O(1).$

We also obtain predictions for the  Z-boson flavor non diagonal decay rates,
$B_{JJ'}= B(Z\to l^-_J+l^+_{J'}) \approx ({\l \l  \over 0.0 1})^2 ({100 GeV \over
\tilde m} )^{2.5} \ 10^{-9} $.  The corresponding CP-odd asymmetries, 
${\cal A}_{JJ'}={B_{JJ'}-B_{J'J}\over B_{JJ'}+B_{J'J} }$,
as induced through interference terms between contributions to the
loop amplitudes in different flavor configurations, 
vary inside the range, $(10^{-1}\ -\ 10^{-3}) \sin \psi $.

\begin{figure} [t]
\centerline{\psfig {figure=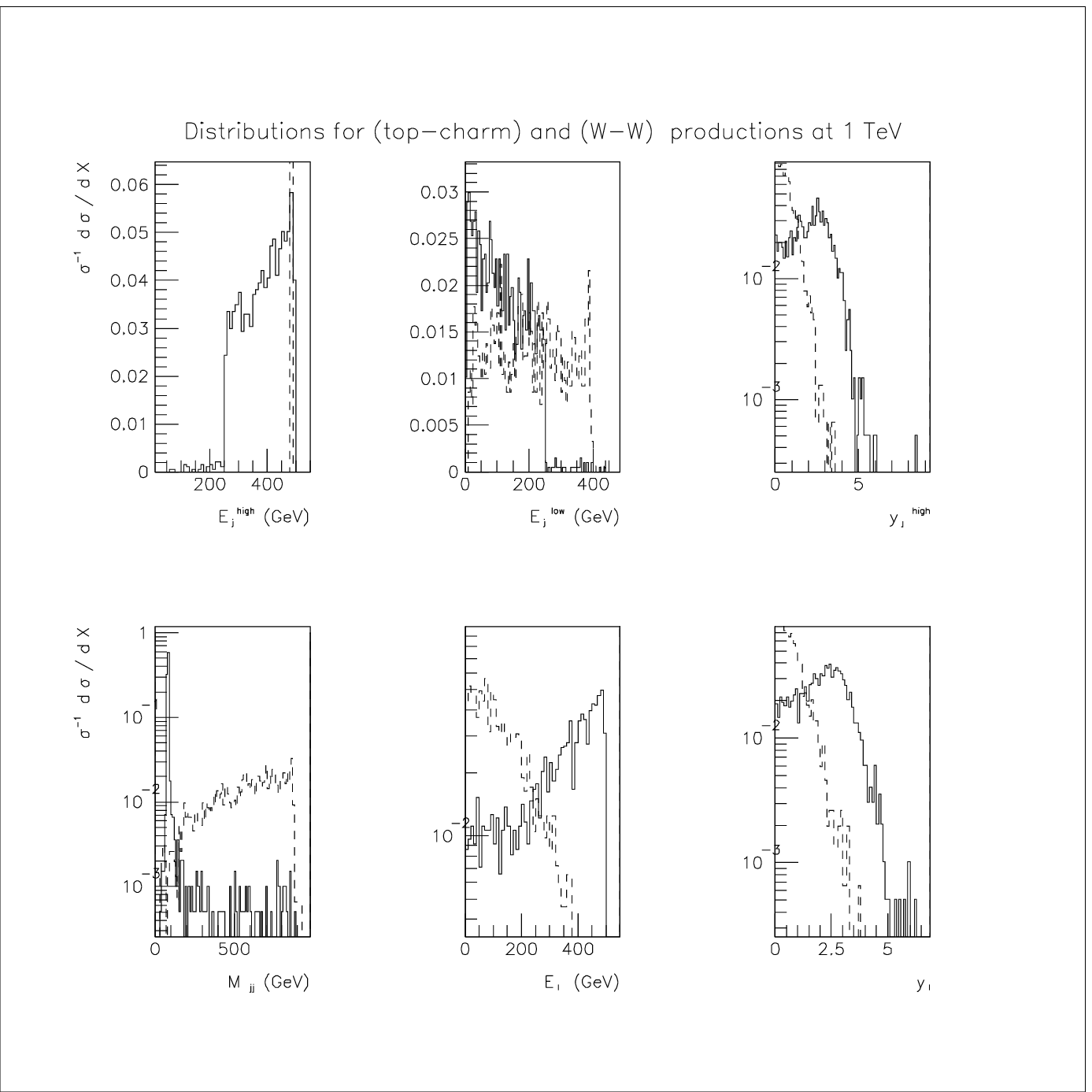,height=9cm,width=12 cm}}
\caption{ Dynamical distributions associated with the signal events,
$l^+l^-\to (t \bar c) +(\bar t c) \to (b l^+ \nu \bar c) +( \bar b
l^-\bar \nu c)$, (dashed line) and the background events, $ l^+l^-\to
W^+W^- \to ( l^+ \nu \bar q q' ) +( ( l^- \bar \nu q\bar q' ) $,
(continuous line) at a center of mass energy, $ s^\ud = 1 TeV$.  
The kinematical variables are the maximum and
minimum energy of the two jets, $ E_j^{high}, \ E_j^ {low}$, the dijet
invariant mass, $ M_{jj} $, and the charged lepton energy, $E_l$, and
rapidity, $ \ y_l = \ud \log {E_l+p_{lz} \over E_l-p_{l z} } $.  }
\label{figtopp}
\end{figure}

The flavor tagging for quarks is experimentally difficult except for
the case of top production. For the top-charm production
reaction,  the final state signal associated with the top semileptonic
decay, $ (b l^+ \nu \bar c) + (\bar b l^- \nu c ), \ [l=e,\ \mu ] $,
consists of an isolated energetic charged lepton, a pair of $ b $
and $ c $ quark hadronic jets and missing energy.  These events have a
clean signature and may easily be distinguished from the standard
model background which arises mainly from the W-boson pair production
reaction, $e^+ e^-\to W^+W^-$.  We consider a set of characteristic
final state kinematical variables. 
 The
distributions, as plotted in Fig.\ref{figtopp}, show marked
differences between signal and background.  The numerically evaluated
efficiencies found for a suitable set of selective cuts are, $ \e_S
\simeq 0.8 $ for the signal and $ \e_B \simeq 3. 10^{-3} $, for the
background. For the interval of  superpartner masses,   $100 < \tilde m = m_{
\tilde d _{kR} } < 1000 GeV $, the sensitivity reach for the relevant RPV 
coupling constant products is,
$ \l '_{12k} \l '_{13k} < (1. \ - \ 5.) \times 10^{-2}
(100 fb /{\cal  L}  ) ^\ud $,  where  $ {\cal L }  $ is the integrated luminosity.

\section{Slepton pair production}

The flavor non-diagonal tree level ($ \nu_i$ exchange) 
amplitudes for the reactions, $l^-+l^+\to
\tilde e^-_{HJ}+\tilde e^+_{H' J'} $, involve the product of RPV
coupling constants, $ \l_{iJ1}^\star \l_{iJ'1} ,$ for L-sleptons and,
$ \l_{i1J}^\star \l_{i1J'} $, for R-sleptons.
The loop amplitudes involve a twofold summation over leptons families
with a quadratic dependence on the RPV coupling constants involving, $
\l ^{'\star }_{Jjk} \l ' _{J'jk} , \  [d_k, \ u_j] \ \l ^{\star }_{Jjk}
\l _{J'jk} ,\   [e_k, \ \nu _j]$, for L-sleptons,  and $ \l _{ijJ} \l
^{\star } _{ijJ'} ,\ [e_j,\ \nu ^c_{i} ] $ for R-sleptons, where the
brackets indicate the fermions circulating in the loops.  Summarizing
briefly our results, \cite{Chemsca} we find that the rates rise
sharply at threshold and, with growing center of mass energy, $s^\ud
$, settle roughly as, $\tilde m^2/s$, to constant values of order, $20
\ - \ 2$ fbarns, in units of $ ({\l ^\star \l \over 0.01} )^2$ or $
({\l ^{' \star } \l ' \over 0.01} )^2$, as one sweeps through the
interval, $\tilde m \in [60, 400]$ GeV for the final sleptons mass.
The predicted flavor non-diagonal CP asymmetries are of
order,\cite{Chemsca} $A_{JJ'}=(10^{-2}-10^{-3}) \sin \psi$, using the
same particular prescription as in the above fermion production case
where the RPV coupling constants products in the loop and tree
amplitudes have the relative phase, $ e^{i\psi }$.
Since the same  quadratic products of coupling constants enter as in the
fermion pair production case,  any observable  contribution
to flavor changing or  CP violation effect in 
fermion pair production would entail a contribution of similar size to
sfermion  pair production.

\end{document}